\documentclass{article}
\usepackage{spconf,amsmath,amssymb,graphicx,booktabs}
\usepackage{multirow}
\usepackage{makecell}
\usepackage[table]{xcolor}

\definecolor{bestblue}{RGB}{218,233,248}
\definecolor{secondorange}{RGB}{255,239,213}
\definecolor{groupgray}{RGB}{244,246,248}

\usepackage{url}
\usepackage[
  colorlinks=false,
  pdfborder={0 0 1},
  linkbordercolor={1 0 0},
  citebordercolor={0 1 0},
  urlbordercolor={0 1 1}
]{hyperref}

\newcommand{\cfo}{\textsc{CFO}}

\title{From Time to Channels: Robust and Efficient Local Flaw Detection in Steel Wire Ropes Using Tri-Axis MFL Signals}

\name{\begin{tabular}{c}
Siyu You$^{1,\dagger}$, Yibo Zhang$^{1,\dagger}$, Shengbo Xu$^{1}$, Yanhui Yang$^{1}$ \\
Huayi Gou$^{1}$, Wen Wu$^{3}$, Bo Du$^{3}$, Fang Xie$^{3}$, Zhiliang Liu$^{1,2\ast}$
\end{tabular}
\thanks{$^{\dagger}$Siyu You and Yibo Zhang contributed equally to this work.
$^{\ast}$Zhiliang Liu is the corresponding author. }}
\address{$^{1}$Glasgow College, University of Electronic Science
and Technology of China\\
$^{2}$School of Mechanical and Electrical Engineering, University of
Electronic Science and Technology of China\\
$^{3}$Sichuan Special Equipment Inspection Institute}

\begin{document}
\ninept
\maketitle

\begin{abstract}
Steel wire ropes (SWRs) are critical load-bearing components whose local flaws (LFs) pose serious safety risks. Magnetic flux leakage (MFL) inspection commonly detects LFs from temporal or axial morphology, which can change with the sensing axis and operating condition. We show instead that LF responses remain localized over neighboring channels of a circular array. This observation motivates a channel-feature-oriented (CFO) detector that removes smooth channel backgrounds, applies circular matched filtering, and fuses spatially co-located tri-axis responses. Experiments performed on real-world equipment show that CFO attains the highest localization performance among three representative baselines, reaching F1@0.5/F1@0.7 scores of 73.9\%/59.5\%. It attains the best temporal localization performance under all four conditions and achieves at least \(19.4\times\) the throughput of the evaluated signal-processing baselines. These results demonstrate that circular channel locality provides a robust LF representation across the evaluated operating conditions.
\end{abstract}

\section{Introduction}

Steel wire ropes (SWRs) are inspected for local flaws (LFs), including broken
wires and corrosion, because such damage can reduce load-bearing capacity
\cite{liu2020review,zhou2019review,kim2018mfl}. Magnetic flux leakage (MFL) inspection
measures the field perturbation produced by a flaw in a magnetized rope
\cite{sun2013mechanism}. A circular Hall array provides circumferential
coverage, and tri-axis sensors capture the axial, radial, and circumferential
field components \cite{yang2025ilf,liu2016circular}. Their responses can vary with the sensing
axis, inspection speed, and motion-induced effects
\cite{piao2021miec,you2025astfo,liu2021comparison}.

Existing MFL signal-processing methods can be broadly divided into two
complementary categories. \emph{Noise-oriented} methods treat shaking,
strand, and background variations as nuisance components and suppress them
using signal statistics, transforms, or spatial correlations. Representative
approaches include shaking-noise elimination based on MFL-image modeling
\cite{ren2021shaking,zhou2021fault,liu2022shaking} and adaptive fast Walsh--Hadamard
transform filtering \cite{huang2024walsh}. Multiscale phase-spectrum
reconstruction has also been explored to enhance weak LF signals
\cite{weaklf}. These methods can improve the
signal-to-noise ratio without assuming a precise defect waveform, but their
effectiveness depends on the separability between defect and nuisance
components, and aggressive suppression may attenuate weak LF responses.

\emph{Defect-feature-oriented} methods instead exploit characteristic MFL
morphologies. TFO constructs an axial target-feature template and enhances its
matched response \cite{pan2024adaptive}. Ensemble methods further combine
axial and channel-direction templates to improve noise robustness
\cite{pan2025ensemble}, while AS-TFO employs multiscale image fusion to
accommodate speed-dependent changes in spatial sampling resolution
\cite{you2025astfo}. By directly enhancing defect-related structures, these
methods can produce concentrated detection responses. Their effectiveness,
however, largely relies on the temporal or axial defect morphology remaining
sufficiently repeatable for template-based detection.

Learning-based detectors extract LF representations directly from labeled
data \cite{pan2023yolo,wang2023cnn, you2025signal}, but their performance depends on the
coverage of the training set. This motivates the search for a more stable LF representation across operating conditions.

Existing channel-direction methods already exploit spatial information, but
their enhancement and localization still depend on axial templates,
interpolation, or speed-dependent time--channel images
\cite{pan2025ensemble,you2025astfo,yang2025ilf}. We instead use the
native circular topology as the primary detection domain. Across axes and
conditions, temporal responses can change in width, phase, and lobe structure,
whereas an LF remains concentrated over a small circular-channel neighborhood.
This motivates our channel-feature-oriented (\cfo) detector,
which does not require a fixed temporal waveform or channel interpolation.
Our contributions are threefold:
\begin{itemize}
  \setlength{\itemsep}{0pt}
  \setlength{\parskip}{0pt}
  \setlength{\parsep}{0pt}
  \item We identify and quantify circular channel locality as a stable LF
  signature across speed and shaking conditions.
  \item We develop a circular channel enhancement front end
  that removes smooth spatial backgrounds and performs matched filtering
  directly on the native sensor array, avoiding fixed temporal templates and
  channel interpolation.
  \item We introduce a circular-neighborhood tri-axis fusion mechanism that
  enforces spatial co-location across the axial, radial, and circumferential
  responses while accommodating small axis-dependent channel offsets.
\end{itemize}
\section{Preliminary Study}
\label{sec:motivation}

\subsection{Data Acquisition}
\label{sec:data}
We used the MFL inspection platform, sensors, and two SWR specimens described
in \cite{weaklf}. One specimen contained four external wire-break LFs involving
one, two, three, and three broken wires, respectively; the other contained
three internal wire-break LFs. An operator manually moved the sensor head
repeatedly back and forth along the rope specimens,
yielding 80 recordings in total. These flaws span representative LF cases across different defect locations and broken-wire severities. In each recording, multiple samples of the physical LFs are obtained under different operating conditions. The platform employs a 32-channel circular Hall-sensor
array. Each sensor measures the
axial ($B_x$), radial ($B_y$), and circumferential ($B_z$) magnetic-field
components at 500~Hz, yielding the tri-axis measurement $x(t,a,c)$ used
throughout this paper. The recordings cover four operating conditions that
span two speed levels and two disturbance settings: $S_1$ and $S_2$
correspond to the low-speed range (0--0.5~m/s) without and with shaking,
respectively, while $S_3$ and $S_4$ correspond to the high-speed range
(0.5--1~m/s) without and with shaking, respectively. The four conditions
contain 80, 110, 120, and 160 annotated LF samples, respectively.

\subsection{Circular Tri-Axis Signal Model}

At sample $t$, sensing axis $a\in\{x,y,z\}$, and circular-array channel
$c\in\{0,\ldots,C-1\}$, let $x(t,a,c)$ denote the preprocessed MFL signal.
We use the following first-order model to characterize a LF response:
\begin{equation}
x(t,a,c)
=
s(t,a)g_a((c-c_0)\bmod C)
+b(t,a,c)+n(t,a,c),
\label{eq:model}
\end{equation}
where $s(t,a)$ denotes the axis- and condition-dependent temporal response,
$g_a$ describes an axis-dependent but spatially localized channel response
centered at $c_0$, $b(t,a,c)$ represents a spatially smooth background, and
$n(t,a,c)$ denotes residual noise. For each complete axis--channel trace, we
subtract a 201-sample local-median baseline and divide by
$1.4826\operatorname{median}|u-\operatorname{median}(u)|$. If this scale is
below $10^{-12}$, we use the trace standard deviation plus $10^{-12}$.
\subsection{Temporal Variability Versus Channel Stability}
\label{sec:ncc}

\begin{figure}[!t]
  \centering
  \includegraphics[width=0.8\linewidth]{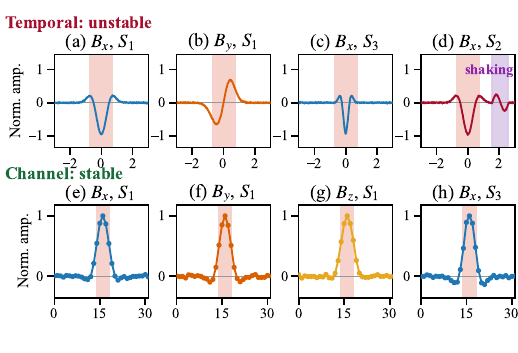}
  \caption{Representative LF responses. Top: normalized waveforms versus
axial position (arbitrary units) across axes and operating changes. Bottom:
normalized channel profiles after circular peak alignment.}
  \label{fig:temporal}
\end{figure}

Fig.~\ref{fig:temporal} illustrates the key observation motivating CFO:
temporal LF morphology varies across sensing axes and operating conditions,
whereas the corresponding responses remain localized over a small
circular-channel neighborhood after suppressing the smooth channel
background.

For each annotated LF interval $I$ and sensing axis $a$, we locate the
peak-magnitude response $(t_a^*,c_a^*)$ in the channel-high-pass signal
$x_{\rm hp}(t,a,c)$ and normalize its polarity. A channel profile is extracted
at $t_a^*$, while a temporal waveform is extracted at $c_a^*$ using the same
fixed 501-sample ($\sim$1-s) peak-centered window for all observations.
Channel profiles are aligned by the circular shift maximizing NCC, whereas
temporal waveforms are aligned by the lag maximizing linear NCC.

The four operating conditions and three sensing axes define 12
condition--axis states. For each source state and array orientation, aligned
responses are averaged to form a reference profile or waveform, which is
compared with responses from the remaining states at the matching array
orientation. For within-condition comparisons, the target LF is excluded
from the corresponding reference. This yields 3378 comparisons in each
domain.

Across these comparisons, the channel-profile NCC is
$75.6\pm16.7\%$ (mean$\pm$standard deviation), compared with
$55.0\pm17.0\%$ for the temporal waveform. The channel/temporal scores are
$76.3\%/61.7\%$ for axis-only changes, $78.6\%/54.1\%$ for condition-only
changes, and $73.8\%/52.9\%$ when both change. These results indicate that,
after suppressing the smooth channel background, the localized channel
structure is more stable across sensing axes and operating conditions than
the corresponding temporal morphology.

\section{Methodology}
\label{sec:method}

\begin{figure*}[!t]
  \centering
  \includegraphics[width=.97\linewidth]{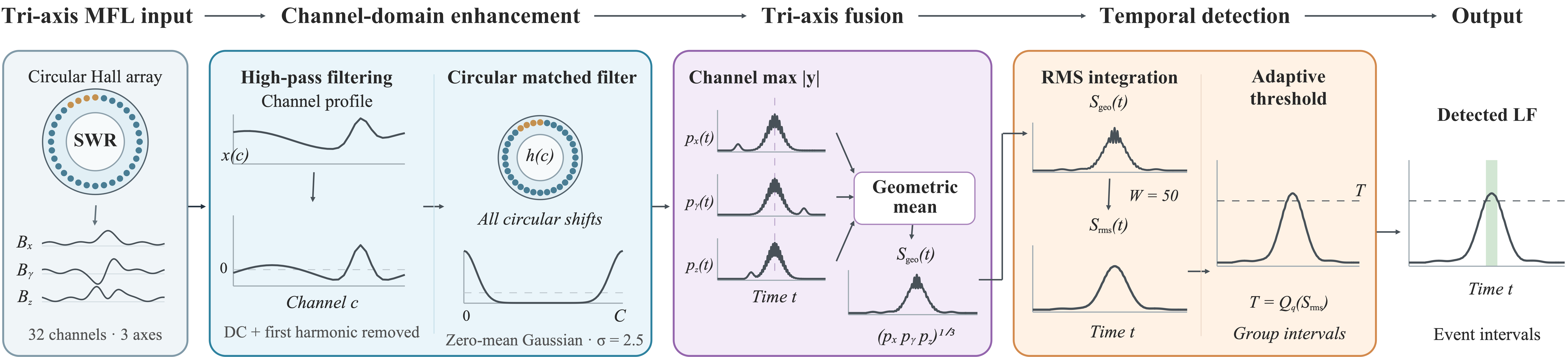}
  \caption{Overview of \cfo{}: circular channel enhancement, spatially
  co-located tri-axis fusion, RMS integration, and adaptive LF sample interval detection.
  Curves are schematic.}
  \label{fig:pipeline}
\end{figure*}

\begin{figure}[!t]
  \centering
  \includegraphics[width=.92\linewidth]{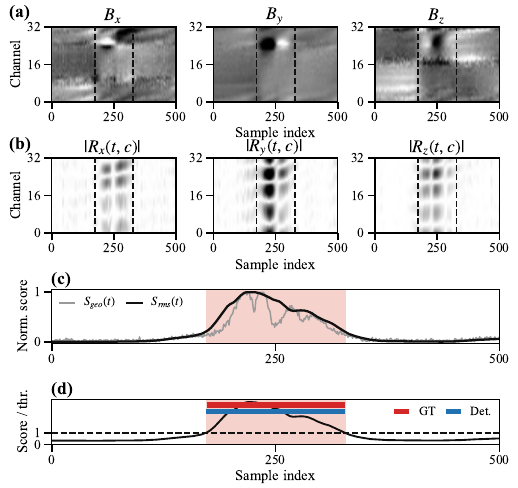}
  \caption{A CFO processing example on a real tri-axis MFL
  recording. (a) Tri-axis MFL signals. (b) Responses after channel-domain enhancement. (c) Tri-axis fused score
  $S_{\rm geo}(t)$ and temporally integrated score $S_{\rm rms}(t)$.
  (d) Adaptive thresholding and final LF localization.
  Dashed lines and shading indicate the annotated LF sample interval.}
  \label{fig:event_example}
\end{figure}

Given the preprocessed tri-axis MFL signal $x(t,a,c)$, where $t$, $a$, and $c$
denote the temporal index, sensing axis, and circular-array channel,
respectively, \cfo{} detects LFs by exploiting their localized
channel-domain responses rather than relying on a fixed temporal waveform.
As illustrated in Fig.~\ref{fig:pipeline}, the framework consists of circular
channel-domain enhancement, tri-axis fusion, and temporal LF sample interval detection. Fig.~\ref{fig:event_example} provides a representative real-world example of
the complete detection process, which is referenced throughout the following
subsections to illustrate the corresponding processing stages.

\subsection{Circular Channel-Domain Enhancement}

A LF typically introduces a localized perturbation over a small
neighborhood of circular-array channels, whereas condition-dependent background
components vary more smoothly along the channel dimension. We therefore first
suppress the low-spatial-frequency channel background.

For each $(t,a)$, a $C$-point DFT is applied along the channel dimension:
\begin{equation}
x_{\rm hp}(t,a,c)=\operatorname{IDFT}_{f}\!\left\{
\operatorname{DFT}_{c}[x(t,a,c)]H(f)\right\},
\label{eq:hp}
\end{equation}
where the circular high-pass mask is
\begin{equation}
H(f)=
\begin{cases}
0, & \min(f,C-f)<K,\\
1, & \text{otherwise}.
\end{cases}
\label{eq:hpmask}
\end{equation}
With $K=2$, the DC and fundamental spatial-frequency components are suppressed,
while localized channel variations are retained.

To enhance the localized channel signature, we construct a zero-mean,
unit-energy circular Gaussian kernel,
\begin{equation}
h(c)=
\frac{
\exp[-d(c)^2/(2\sigma^2)]-\bar h
}{
\left\|
\exp[-d(c)^2/(2\sigma^2)]-\bar h
\right\|_2
},
\quad
d(c)=\min(c,C-c),
\label{eq:kernel}
\end{equation}
where $\bar h$ denotes the channel-wise mean of the Gaussian profile.
Circular matched filtering is implemented efficiently in the DFT domain:
\begin{equation}
y(t,a,c)=
\mathrm{IDFT}\left\{
\mathrm{DFT}[x_{\rm hp}(t,a,\cdot)]
\,
\mathrm{DFT}[h]^*
\right\}.
\label{eq:match}
\end{equation}
This operation evaluates the localized defect pattern at all circumferential
positions while preserving the native circular topology of the $C$ sensor
channels. We use $\sigma=2.5$, corresponding to the observed localized
channel extent.

\subsection{Tri-Axis Fusion}

Let $d_C(c,c')=\min(|c-c'|,C-|c-c'|)$ be circular channel distance. We first
allow a radius-$r$ alignment tolerance for each axis and then fuse co-located
responses; $r$ is set to 2:
\begin{equation}
S_{\rm geo}(t)=\max_c\left[\prod_{a\in\{x,y,z\}}
\max_{c':d_C(c,c')\le r}|y(t,a,c')|\right]^{1/3}.
\label{eq:fusion}
\end{equation}
This score emphasizes tri-axis responses occurring at the same time and within
a small circular neighborhood; it does not require the three axis maxima to
fall on the same channel.

\subsection{Temporal Integration and Adaptive LF Sample Interval Detection}

The fused score may contain short fluctuations. We use a centered RMS window
$\mathcal W_t=\{\tau:\max(1,t-25)\le\tau\le\min(N,t+25)\}$:
\begin{equation}
S_{\rm rms}(t)=
\sqrt{\frac{1}{|\mathcal W_t|}\sum_{\tau\in\mathcal W_t}
S_{\rm geo}^{2}(\tau)},
\label{eq:rms}
\end{equation}
which uses 51 samples away from recording boundaries and clips the window at
the two ends.

An adaptive per-recording threshold is defined as the $q$-quantile of
the integrated score:
\begin{equation}
T=\mathrm{Quantile}_{q}\left(\left\{S_{\rm rms}(t)\right\}_{t=1}^{N}\right),
\label{eq:threshold}
\end{equation}
where $q=0.925$ is fixed for every condition and method; labels are not used to
estimate a recording's threshold. Contiguous samples satisfying
\begin{equation}
S_{\rm rms}(t)>T
\end{equation}
form candidate LF sample intervals, which must last at least 40 samples. Their boundaries
are refined on a centered 31-sample RMS curve at 15\% of the peak-to-local-
background range, with a 20-sample search pad. Refinement is accepted only for
candidates no longer than 150 samples and when it reduces duration by at least
20\%; overlapping ground-truth LF sample intervals are merged before evaluation.
\section{Experimental Evaluation}
\label{sec:experiments}

\subsection{Experimental Setup}

All recordings use the platform in Sec.~\ref{sec:data}. Window metrics use
500-sample windows with a 250-sample stride; an incomplete tail is discarded.
A window is positive if it overlaps an LF annotation, and its anomaly score is
the maximum $S_{\rm rms}(t)$ in that window. This produces 1,632 windows.

\subsection{Evaluation Protocol and Baselines}

Window AUROC/AUPRC are computed from pooled window scores, while LF interval detection is performed once per complete recording. For each tIoU threshold, predicted LF sample intervals
are processed in descending peak-score order and greedily matched to the
unmatched ground-truth LF sample interval with the largest tIoU, yielding a one-to-one
assignment. We report precision, recall, and F1 at tIoU$=0.5$, and F1 at 0.7.

We compare \cfo{} with TFO~\cite{pan2024adaptive},
Ensemble~\cite{pan2025ensemble}, and AS-TFO~\cite{you2025astfo}. TFO and
AS-TFO sum their three axis responses, while Ensemble takes the maximum; all
three therefore receive the same tri-axis input available to CFO. The baseline
front ends follow the cited implementations. Four pilot recordings, one from
each of S1--S4, were used to tune the method-specific front-end parameters
within the tested parameter ranges. The pilot recordings were excluded from
the final evaluation, and the selected parameters were then fixed for all
remaining recordings and operating conditions. To isolate the effect of the
front ends, we replace their native single-axis localization gates with the
shared RMS integration, $q=0.925$ threshold, duration filter, and boundary
refinement above. The comparison therefore evaluates tuned front ends under a
common LF sample interval generation procedure, without condition-specific or
recording-specific retuning.

Overall interval metrics are micro-aggregated over recordings, while AUROC/AUPRC use all 1,632 windows. Front-end throughput is measured on an Intel Core i5-14500 after common preprocessing.

\subsection{Overall and Condition-Wise Results}

Table~\ref{tab:condition_results} summarizes both condition-wise and overall
performance. From the results, CFO achieves the best overall AUROC/AUPRC (96.5\%/91.4\%) and F1@0.5/F1@0.7 (73.9\%/59.5\%). Compared with Ensemble, the strongest baseline in interval localization, CFO improves F1@0.5 and F1@0.7 by 12.0 and 24.2 percentage points, respectively. The improvement is also consistent across the four operating conditions.
\begin{table}[!t]
\centering
\caption{Condition-wise comparison of training-free LF detectors. Best results under each condition are shown in bold.
For the overall results, blue and orange indicate the best and
second-best performance, respectively.}
\label{tab:condition_results}

\setlength{\tabcolsep}{2.8pt}
\renewcommand{\arraystretch}{1.05}

\resizebox{0.82\columnwidth}{!}{%
\begin{tabular}{llcccccc}
\toprule
Cond. & Method & AUROC & AUPRC & P@0.5 & R@0.5 & F1@0.5 & F1@0.7 \\
\midrule

\multirow{4}{*}{$S_1$}
& TFO
& 96.6 & 94.7 & 48.5 & 60.0 & 53.6 & 34.6 \\

& Ensemble
& 96.7 & 95.0 & 48.5 & 60.0 & 53.6 & 38.0 \\

& AS-TFO
& 96.3 & 93.7 & 70.4 & 62.5 & 66.2 & 34.4 \\

& \textbf{CFO}
& \textbf{97.6}
& \textbf{96.6}
& \textbf{69.1}
& \textbf{70.0}
& \textbf{69.6}
& \textbf{60.9} \\

\midrule

\multirow{4}{*}{$S_2$}
& TFO
& 93.9
& 80.3
& 70.9
& 66.4
& 68.6
& 42.3 \\

& Ensemble
& 92.4
& 80.4
& 72.0
& 71.0
& 71.7
& 36.1 \\

& AS-TFO
& 89.4
& 71.8
& 24.4
& 45.5
& 31.7
& 12.7 \\

& \textbf{CFO}
& \textbf{94.3}
& \textbf{80.6}
& \textbf{72.5}
& \textbf{71.8}
& \textbf{72.1}
& \textbf{46.6} \\

\midrule

\multirow{4}{*}{$S_3$}
& TFO
& 96.9
& 86.2
& 66.7
& 73.3
& 69.8
& 44.4 \\

& Ensemble
& 97.0
& 84.0
& 74.2
& 76.7
& 75.4
& 45.9 \\

& AS-TFO
& 97.0
& 86.6
& 72.3
& 78.3
& 75.2
& 30.4 \\

& \textbf{CFO}
& \textbf{97.6}
& \textbf{90.3}
& \textbf{76.3}
& \textbf{96.7}
& \textbf{85.3}
& \textbf{80.9} \\

\midrule

\multirow{4}{*}{$S_4$}
& TFO
& 93.3
& 87.5
& 33.3
& 40.2
& 36.5
& 16.7 \\

& Ensemble
& 96.4
& 91.9
& 47.9
& 51.7
& 49.7
& 28.7 \\

& AS-TFO
& 95.4
& 90.8
& 45.6
& 42.5
& 44.0
& 21.4 \\

& \textbf{CFO}
& \textbf{97.1}
& \textbf{94.3}
& \textbf{60.2}
& \textbf{67.8}
& \textbf{63.8}
& \textbf{47.6} \\

\midrule

\multirow{4}{*}{\textbf{Overall}}
& TFO
& 94.9
& \cellcolor{secondorange}\textbf{88.2}
& 51.8
& 58.1
& 54.8
& 32.3 \\

& Ensemble
& \cellcolor{secondorange}\textbf{95.6}
& 86.8
& \cellcolor{secondorange}\textbf{59.8}
& \cellcolor{secondorange}\textbf{64.3}
& \cellcolor{secondorange}\textbf{61.9}
& \cellcolor{secondorange}\textbf{35.3} \\

& AS-TFO
& 95.2
& 86.9
& 47.2
& 55.7
& 51.1
& 22.8 \\

& \textbf{CFO}
& \cellcolor{bestblue}\textbf{96.5}
& \cellcolor{bestblue}\textbf{91.4}
& \cellcolor{bestblue}\textbf{70.4}
& \cellcolor{bestblue}\textbf{77.8}
& \cellcolor{bestblue}\textbf{73.9}
& \cellcolor{bestblue}\textbf{59.5} \\

\bottomrule
\end{tabular}%
}

\end{table}

\subsection{Computational Efficiency}

Let $N$, $A$, and $C$ denote the temporal length, number of sensing axes, and
number of circular channels, while $k_t$, $k_c$, and $L$ denote the temporal
template size, channel template size, and number of pyramid levels. CFO uses
short DFTs along the circular-channel dimension with linear-complexity
fixed-radius tri-axis fusion, yielding a dominant complexity of
$\mathcal{O}(ANC\log C)$.

\begin{table}[!t]
\centering
\caption{Theoretical complexity and implementation-controlled front-end
throughput. I/O, preprocessing, and post-processing
are excluded. Relative speed is normalized to TFO.}
\label{tab:efficiency}

\setlength{\tabcolsep}{3.8pt}
\renewcommand{\arraystretch}{1.05}

\resizebox{0.82\columnwidth}{!}{%
\begin{tabular}{lccc}
\toprule
Method & Dominant Complexity & Throughput (windows/s) & Rel. Speed \\
\midrule

TFO
& $\mathcal{O}(ANCk_t)$
& 40.3
& $1.0\times$ \\

Ensemble
& $\mathcal{O}\!\left(ANC(k_t+k_c)\right)$
& 22.3
& $0.55\times$ \\

AS-TFO
& $\mathcal{O}(ALNCk_tk_c)$
& 24.2
& $0.60\times$ \\

\textbf{CFO}
& $\mathbf{\mathcal{O}(ANC\log C)}$
& \textbf{779.5}
& \textbf{19.35}$\times$ \\

\bottomrule
\end{tabular}%
}
\end{table}

As shown in Table~\ref{tab:efficiency}, CFO achieves 779.5 windows/s,
corresponding to $19.35\times$ the throughput of TFO.

\subsection{Ablation and Sensitivity Analysis}
\label{sec:ablation}

We further evaluate the contributions of the key CFO components and the
sensitivity to major design parameters. All experiments follow the same protocol as the main evaluation, with only one component or
parameter changed at a time.

\begin{table}[!t]
\centering
\caption{Ablation and parameter sensitivity of CFO. }
\label{tab:ablation_sensitivity}

\setlength{\tabcolsep}{2.6pt}
\renewcommand{\arraystretch}{1.06}

\resizebox{0.82\columnwidth}{!}{%
\begin{tabular}{l|c|cc||l|c|cc}
\toprule

\multicolumn{8}{c}{\textit{Module Ablation}} \\
\midrule

\multicolumn{6}{l}{\textbf{Configuration}}
& \textbf{F1@0.5} & \textbf{F1@0.7} \\

\cmidrule(lr){1-8}

\multicolumn{6}{l}{\textbf{Full CFO}}
& \textbf{73.9} & \textbf{59.5} \\

\multicolumn{6}{l}{w/o circular high-pass}
& 37.8 & 23.9 \\

\multicolumn{6}{l}{w/o matched filtering}
& 69.0 & 51.0 \\

\multicolumn{6}{l}{Axis-max fusion}
& 71.2 & 56.7 \\

\midrule

\multicolumn{8}{c}{\textit{Parameter Sensitivity}} \\
\midrule

\textbf{Parameter} & \textbf{Value} & \textbf{F1@0.5} & \textbf{F1@0.7}
& \textbf{Parameter} & \textbf{Value} & \textbf{F1@0.5} & \textbf{F1@0.7} \\
\midrule

\multirow{3}{*}{$\sigma$}
& 1.5
& 72.1
& 57.8
& \multirow{3}{*}{$K$}
& 1
& 37.8
& 23.9 \\

& \textbf{2.5}
& \textbf{73.9}
& \textbf{59.5}
&
& \textbf{2}
& \textbf{73.9}
& \textbf{59.5} \\

& 3.5
& 70.8
& 57.7
&
& 3
& 69.8
& 55.0 \\

\cline{1-4}
\cline{5-8}

\multirow{3}{*}{RMS length}
& 31
& 71.1
& 56.5
& \multirow{3}{*}{$q$}
& 0.900
& 63.2
& 51.4 \\

& \textbf{51}
& \textbf{73.9}
& \textbf{59.5}
&
& \textbf{0.925}
& \textbf{73.9}
& \textbf{59.5} \\

& 71
& 72.5
& 58.6
&
& 0.950
& 67.4
& 47.5 \\

\bottomrule
\end{tabular}%
}
\end{table}
Circular high-pass filtering has the largest effect in the ablation study:
removing it reduces F1@0.5 by 36.1 points. Since the matched kernel is
zero-mean, $K=1$ removes only the DC component and is equivalent here to
omitting the additional high-pass stage, whereas $K=2$ additionally suppresses
the first spatial harmonic. Matched filtering and spatially co-located tri-axis
fusion further improve localization performance.

Table~\ref{tab:ablation_sensitivity} also shows that CFO is relatively stable around the
selected parameter values. The choices $K=2$, $\sigma=2.5$, and $r=2$ are
consistent with the assumed smooth channel background, localized channel
response, and small inter-axis channel offsets. Among the tested temporal
integration settings, the 51-sample RMS window gives the highest overall
localization performance, with F1@0.5/F1@0.7 scores of 73.9\%/59.5\%.
Using 31 or 71 samples leads to only moderate degradation. The results also show
a clear optimum around $q=0.925$ among the evaluated threshold values.

\section{Conclusion}

We presented CFO, an LF detector that exploits stable locality over circular sensor channels instead of variable temporal morphology. Across four speed and shaking conditions, CFO consistently outperformed representative baselines in LF localization while retaining high computational efficiency. These results demonstrate the effectiveness of circular channel locality for robust and efficient tri-axis MFL-based LF detection.
\section{Acknowledgement}

This work was supported by the Science and Technology Program of the State Administration for Market Regulation (2025MK128).
\bibliographystyle{IEEEbib}
\bibliography{reference}
\end{document}